\documentclass{elsart}
\begin{document}
\newcommand{\dsp}{\displaystyle}
\begin{frontmatter}
\title{Numerical modelling of the quantum-tail effect on fusion 
rates at low energy\thanksref{RFBRmiur}}
\author[caUNI,caINFN]{M. Coraddu},
\author[caUNI,caINFN]{G. Mezzorani},
\author[TRINITI]{Yu.~V. Petrushevich}
\author[toPOLI,caINFN]{P. Quarati},
\author[TRINITI]{A.~N. Starostin}
\corauth[corr]{Corresponding author.}
\ead{A.Starostin@relcom.ru}
\address[caUNI]{Physics Dept.,  Univ. Cagliari, 
I-09042 Monserrato, Italy}
\address[caINFN]{I.N.F.N. Cagliari, 
I-09042 Monserrato, Italy}
\address[TRINITI]{State Research Center of Russian Federation 
Troitsk institute for innovation and fusion research,
Center for Theoretical Physics and Computational Mathematics,
Troitsk, 142190 Moscow reg., Russia}
\address[toPOLI]{Physics Dept., Politecnico Torino, 
I-10125 Torino, Italy}
\thanks[RFBRmiur]{This 
work was partially supported by RFBR grant
No. 02-02-16758-a, grant of the president of 
Russian Federation No. NS-1257.2003.2 and
MIUR (Ministero dell'Istruzione,
del\-l'Uni\-ver\-si\-t\`a e della Ricerca) under MIUR-PRIN-2003 project 
``Theoretical Physics of the Nucleus and the Many-Body Systems''.
}

\begin{abstract}
Results of numerical simulations of fusion rate d(d,p)t, for low-energy 
deuteron beam, colliding with deuterated metallic matrix 
(Raiola \cite{Raiola1,Raiola2}) confirm analytical
estimates given in Ref.~\cite{CoradduA:next2003} 
(M. Coraddu {\em et al.}, this issue), 
taking into account quantum tails in the momentum distribution 
function of target particles, and predict an enhanced 
astrophysical factor in the  1 keV region in qualitative
agreement with experiments.
\end{abstract}
\end{frontmatter}

\section{Introduction}
Significant divergence from theoretical predictions of
non-resonant fusion cross-section  at low energies of
incident deuteron particles has recently been observed in 
experimental works \cite{Raiola1,Raiola2}.
At energies of the charged deuteron beam less than 5 keV,
colliding with deuterated metallic matrix, a great
enhancement of the fusion cross section takes place,
compared to theoretical evaluations.  
The hypothesis of ions interacting via screened potential
inside the metallic target was proposed in
\cite{Raiola1}, but an unrealistically large 
screening potential was necessary
to obtain reasonable agreement with the observed
data. It is obvious that problems arising from
experimental results need  refining
the theoretical models actually used.

As it was shown in papers  \cite{starostin00,starostin02}, the  role
of quantum corrections to the particle momentum distribution 
is quite important. A significant deviation from Maxwellian 
distribution function appears.
At large values of momentum, due to quantum corrections,
the tail of the distribution has a power-law asymptotic 
behavior instead of exponential, 
resulting in modified reaction rates.
In particular, this effect leads to non-exponential temperature
dependence of inelastic process rates at relatively 
low temperatures and high densities.

In this paper we present results of numerical calculations 
of reaction rates for conditions of the interacting 
particles in the  beam and in the target such as those 
of the experiments \cite{Raiola1,Raiola2}.

\section{Numerical modeling of reaction rates for experimental conditions}
 \label{sec:1}
The state of the system  defined by the generalized distribution function over
energy and momentum  $F(E,\vec{p})$, may be presented in the
factorized form:
$F(E,\varepsilon)=n(E)a(E-\varepsilon)$, where $\varepsilon=p^2/2m$
is the particle kinetic energy.
The reaction-rate constants of the inelastic process between two
particles, named ``a'' and ``b'', may be presented in a more general form
by the integral (see  Ref.~\cite{Starostin:next2003}, 
A. Starostin et al.,this issue)
\begin{equation}
  \label{integ}
\begin{array}{c}
  N_aN_b K_{ab}= 8\pi C
          \int\limits_{-\infty}^{\infty} dE_a
          \int\limits d\vec{p}_a
          \int\limits_{-\infty}^{\infty} dE_b
          \int\limits d\vec{p}
          \int d\vec{q}
          \times \\
        n(E_a)\bigl(1-n(E_a+Q_a-\omega)\bigr) a_a(E_a-\varepsilon_a) \times \\
        n(E_b)\bigl(1-n(E_b+\omega+Q_b)\bigr) a_b(E_b-\varepsilon_b) \times \\
        a'_a(E_a+Q_a-\omega-\varepsilon_{\vec{p}_a-\vec{q}})
          a'_b(E_b+\omega+Q_b-\varepsilon_{\vec{p}_b+\vec{q}})
          \sigma(\varepsilon_p)\dsp{\sqrt{\frac{2\varepsilon_p}{M}}},
\end{array}
\end{equation}
where $E_a$ , $p_a$ are the  energy and momentum of the ``a'' particle,
$\varepsilon_p=p^2/2M$ is the  relative kinetic energy in the center
of mass, $M$ is the reduced mass of colliding particles, $C$ is a
normalization constant, defined from comparison of the expression calculated
by (\ref{integ}) and known results at high temperature and low
density. We have:
$$
  \vec{p}=\frac{m_b\vec{p_a}-m_a\vec{p_b}}{m_a+m_b} \; \; \; .
$$
The expression for the population number $n(E)$ depends 
on the statistical distribution of the system. For the purpose
of this work we must consider that deuterons, which are bosons,
have the distribution
$$
  n(E)=(\exp[(E-\mu)/T]-1)^{-1}
$$
where $\mu$ is chemical potential. For non-ideal plasma, 
the spectral dependence of the  distribution function, 
defined by Lorentzian profile, is:
 \begin{equation}
  \label{width}
   a(E,\vec{p}) = \frac{\gamma_i(E,\vec{p})}{\pi
         \left[ \left(E-\varepsilon_p-\Delta(E,\vec{p}) \right)^2
                + \gamma_i^2(E,\vec{p}) \right]} \; \; \; . 
   \label{eq:ai1}
\end{equation}
In Eq.~(\ref{width}) $\gamma$ is the width, $\Delta$ stands for the
energy shift due to atom-matter interaction.
The line width is given by
$\gamma_a=N\hbar\sigma_a V_a$; 
where $\sigma_a=\pi e^4/\varepsilon_a^2$, 
$V_a=\sqrt{2\varepsilon_a/m_a}$, $N$ is the
concentration of scattering centers. For ideal plasma conditions,
i.e. when the density decreases, the width
$\gamma(E,\varepsilon_p)\to 0$ , the function $a(E,\vec{p})$
becomes a  delta-function.
The cross-section dependence on kinetic energy 
may be given in the form:
 \begin{equation}
   \label{ther}
   \sigma_0(\varepsilon_p)=
\frac{S(\varepsilon_p)}{\varepsilon_p}\exp(-2\pi\eta(\varepsilon_p)),\;\;\;
   \eta(\varepsilon_p)=\frac{Z_1Z_2 e^2}{\hbar}
  \sqrt{\frac{M}{2\varepsilon_p}}
 \end{equation}
 where $\eta(\varepsilon_p)$ is the  Sommerfeld parameter.
The astrophysical factor $S$ is weakly dependent on the kinetic energy.

The influence of the screening potential  $U_e$
on the reaction rate,
due to the effect of metallic electrons, may be taken into account by
adding $U_e$  to the collision energy:
\begin{equation}
 \label{Poten}
 \sigma(\varepsilon_p)=\sigma_0(\varepsilon_p+U_e) \; \; \; .
\end{equation}
Numerical calculations of the reaction rates were performed in
accordance with the above model, for conditions close to
the experimental ones: target particles concentration
$N_a=5 \cdot 10^{23}cm^{-3}$, and interacting particle 
masses: $m_a=m_b=2$ amu. 
Fusion reactions  were considered between target
particles of kind ``a''  and beam particles of kind ``b''.
In  the expression of the width we used as  
concentration $N$ the concentration of the 
scattering ions in the metallic matrix. 

Taking into account high dimension, the computation of 
the integral (\ref{integ})
can be performed using Monte Carlo method. The kinetic
energy distribution of target particles was taken at
temperature $T=2.44\cdot10^{-2}eV$, while the ``b'' particles
were taken with the beam energies. It is interesting to note, 
that for the case of a beam of mono-energetic particles the 
reaction rate computation can be reduced from expression
 (\ref{integ}) to the more simple integral:
\begin{equation}
   \label{integ11}
   N_a K^{'}=C\int_0^{\infty}dE_a\int d\vec{p}
   n_a(E_a)a(E_a-\varepsilon_p,\varepsilon_p)
   \dsp{\sqrt{\frac{2\varepsilon_p}{M}}}
\sigma(\varepsilon_p)d\varepsilon_p \; \; \; .
\end{equation}
Within the framework of such model, for the case of ideal plasma, when we can
neglect the wings of the Lorentzian profile in the 
integral (\ref{integ11}), the expression for the reaction rate
can be further simplified:
 \begin{equation}
  \label{integ1}
  \begin{array}{c}
  \dsp{N_a K_{2}=N_a K^{'}=C \int_0^{\infty}dE_a\int d\vec{p}
   \int n_a(E_a)\delta(E_a-\varepsilon_p)
   \dsp{\sqrt{\frac{2\varepsilon_p}{M}}}\sigma(\varepsilon_p)d\varepsilon_p
   \sim} \\
   \dsp{\sim \int_0^{\infty} n_a(\varepsilon_p)\varepsilon_p
   \sigma(\varepsilon_p)d\varepsilon_p} \; \; \; .  \\   
 \end{array}
 \end{equation}
The influence of the distribution wings on the reaction rate value
can be obtained by comparison of the computation results of the two
expressions (\ref{integ}) and (\ref{integ11}). We can also
compare such results with the calculated reaction rate  $K_1=\sigma V$,  
using expression (\ref{ther}).
After such comparison it is possible to estimate the astrophysical
factor $S$  and the deviation of theoretical predictions from
experimental data.

To estimate the influence of momentum distribution tails
on the reaction rate  and the  difference
with the Maxwellian case, it is necessary to take into account the finite
width of the Lorentzian profile (\ref{width}).   
As was shown in \cite{starostin02} and in \cite{Starostin:next2003},
the main result of quantum corrections is that the momentum
distribution function has asymptotically a power-law tail:
\begin{equation}
 f(\varepsilon_p)=C^{'}\int_0^{\infty}dE_a
 a(E_a-\varepsilon_p,\varepsilon_p)\sim
 \exp(-\varepsilon_p/T)+C_a(T)/\varepsilon_p^4 \; \; \; .
\end{equation}
Using such decomposition in equation (\ref{integ11}), 
it is possible to calculate the reaction
rate, taking into account non-Maxwellian distribution function:
\begin{equation}
\label{eq:k3}
 K_3=C_3\int_0^{\infty}d\varepsilon_a
 f(\varepsilon_a)\sqrt{
\frac{\varepsilon_p\varepsilon_a}{M}}\sigma(\varepsilon_p) \; \; \; .
\end{equation}
 The results of such calculations are shown in the table~\ref{tab:1}.

\begin{table}
\caption{Comparison of the reaction rates $\langle \sigma v \rangle$ 
as function of the energy of the beam using the general 
expression of Eq.~(\ref{integ}), $K$, and the three models 
in Eq.~(\ref{ther}), $K_1$, in  Eq.~(\ref{integ11}), $K_2$, and 
in  Eq.~(\ref{eq:k3}), $K_3$.
\label{tab:1}
}
 \begin{center}
 \begin{tabular}{c c c c c c}
\hline \hline
   $E_a$ [keV] & $K_1$ & $K_2$ & $K_3$ & $K$ & $K/K_1$ \\
\hline
  15 & 4.381E+04 &  4.045E+04  & 7.393E+04 &  4.38E+04 & 1.00E+00 \\
  10 & 4.073E+03 &  3.762E+03 &  6.877E+03 &  4.11E+03  &  1.01E+00\\
  5  & 1.711E+01 &  1.580E+01 &  2.892E+01 &  1.77E+01  &  1.03E+00\\
 2 & 2.615E-04 &  2.421E-04 &  4.487E-04 &  2.85E-04  &  1.09E+00\\
 1.8 & 5.038E-05 &  7.223E-05 &  1.344E-04 &  5.62E-05 & 1.12E+00\\
 1.5 & 2.339E-06 &  3.850E-06 &  7.343E-06  & 3.34E-06 &
 1.43E+00\\
 1.2 & 3.613E-08 &  7.474E-08 &  2.265E-07 &  7.84E-07  &
 2.17E+01\\
 1 & 8.252E-10 &   7.711E-10 &  5.678E-08 &  2.82E-07 &
 3.42E+02\\
 \hline \hline
\end{tabular}
\end{center}
\end{table}

The reaction rate constants,
calculated  by different models, agree among themselves at
beam energies above  $2$\/ keV. Decreasing energy in the
range between $2$\/ keV and $1$\/ keV, we have found that
$K_1$\/ and $K_2$\/ have still close values, while the 
rate constant $K$\/ is much larger.  It is interesting to note 
that the constants $K_3$\/ and $K$\/ have relatively close values.
Thus, it is
possible to conclude, that, for correct estimations of the rate
constants, it is quite possible to use the expressions shown in 
(\ref{integ11}). The results of these calculations show that the
wings of the momentum distribution are very important for a correct
evaluations of the reaction rates.  
In the last column of table~\ref{tab:1} we presented the factor 
which characterizes the deviation of the rate
in cause of  non-ideal plasmas.

It is interesting also to estimate the role of the screening effect
on the reaction rate and to compare it to the considered mechanisms.
For that purpose the calculations were performed within the framework
of the proposed model, but with addition of a 
screening potential $U_e =28$\/ eV.
Its influence was taken into account in accordance with expression
(\ref{Poten}). Such value of the potential seems realistic for 
the experimental conditions \cite{Raiola1}.  The results
of calculations are presented in the table~\ref{tab:2}.

\begin{table}
\caption{Same as table~\ref{tab:1} taking into account the
screening effect according to Eq~(\ref{Poten}) with $U_e =28$ eV.
\label{tab:2}
}
\begin{center}
 \begin{tabular}{c c c c c }
 \hline\hline
  $E_a$ [keV] & $K_1$ &   $K_2$   &   $K_3$ & $K_3/ K_1$\\
 \hline
  15 &  4.474E+04 &  4.133E+04  & 7.552E+04 &  1.6879937\\
  10 & 4.237E+03  & 3.914E+03 &  7.152E+03  & 1.6880552 \\
   5 &  1.911E+01 & 1.766E+01 & 3.232E+01  & 1.69143\\
   2 &  4.059E-04 &  5.370E-04 & 9.931E-04 & 2.4466201\\
  1.8 &  8.433E-05 &  1.184E-04 &   2.197E-04  & 2.6056397 \\
  1.5 & 4.605E-06 & 7.336E-06 &   1.383E-05 &   3.0034846 \\
  1.2 & 9.315E-08 & 1.819E-07 & 4.320E-07 & 4.638S3085\\
  1 & 2.863E-09 & 6.951E-09 & 7.386E-08 & 25.801192 \\
 \hline\hline
\end{tabular}
\end{center}
\end{table}

Here we have not presented the results of the more general 
computations using equation
(\ref{integ}), because we have obtained rather correct estimates using
model (\ref{integ11}).  We find a weak influence of the
screening effect, using the reasonable value of the potential, which
agrees with the results of \cite{Raiola1,Raiola2}. 
Taking into account quantum corrections, the theoretical 
evaluations of the rates increase,
in the low-energy range at  $1$-$2$\/ keV and less, if compared
to the rates evaluated without the quantum effect.


\begin{thebibliography}{99}
\bibitem{Raiola1} 
F.~Raiola {\it et al.},
Phys.\ Lett.\ {\bf B 547} (2002) 193.
%
\bibitem{Raiola2}
F.~Raiola {\it et al.},
Eur.\ Phys.\ J. {\bf A 13} (2002) 377.
%
\bibitem{CoradduA:next2003}
M.~Coraddu, M.~Lissia, G.~Mezzorani, Yu.~V.~Petrushevich,
P.~Quarati, and A.~N.~Starostin,
``Quantum-tail effect in low energy d+d reaction 
in deuterated metals,''
Physica A, this issue. 
%
\bibitem{starostin00}
A.~N.~Starostin, V.~I.~Savchenko, and N.~J.~Fisch,
Phys. Lett. A {\bf 274} (2000) 64.
%
\bibitem{starostin02}
A.~N.~Starostin, A.~B.~Mironov, N.~L.~Aleksandrov, 
J.~N.~Fisch, and R.~M.~Kulsrud,
Physica A {\bf 305} (2002) 287.
%
\bibitem{Starostin:next2003}
A.~N.~Starostin, A.~G.~Leonov, Yu.~V.~Petrushevich,
Vl.~K.~Roerich,
``Distribution function, reaction rates, and spectral line
profile,''
Physica A, this issue. 
\end{thebibliography}
\end{document}